\documentclass[aps,prl,preprint,groupedaddress]{revtex4}
\usepackage{graphicx}
\pdfoutput=1

\begin{document}


\title{CdZnTe room-temperature semiconductor operation in liquid scintillator}


\author{D.Y. Stewart}
\email[]{d.y.stewart@warwick.ac.uk}
\author{Y. Ramachers}
\email[]{y.a.ramachers@warwick.ac.uk}
\affiliation{Dept. of Physics, University of Warwick, Coventry CV4 7AL, UK}


\date{}

\begin{abstract}
We demonstrate the first operation of CdZnTe room-temperature detectors in a
liquid scintillator environment. This work follows conceptually the
Heusser-type detector method of operating HPGe detectors in liquid nitrogen
and liquid argon but instead for a far more practical room-temperature
ensemble with the aim of achieving ultra-low background levels for radiation
detection. 
\end{abstract}


\maketitle

\section{Introduction}
The motivation to research CdZnTe (CZT) detectors originates from
participation in the COBRA experiment \cite{cobra}, a proposed massive (several 
hundred kg) array of CZT crystals for double-beta decay research \cite{review}. 
Taking into account a typical 
mass of just a few grams for each crystal, several tens of thousands of
crystals will eventually have to be operated reliably over several
years at an extremely low level of background, under $10^{-3}$ counts per keV,
kg and year in the signal region. 
\par
A relatively new detector concept to achieve such extremes of low background
levels for semiconductor detectors has been proposed by Heusser \cite{heusser}
and later applied to large-scale double-beta decay projects
\cite{genius}\cite{gerda}. Initially such a Heusser-type detector would use a
passive but ultra-clean environment like liquid nitrogen to cool and shield
HPGe detectors. Recently, the immersion of HPGe's in an active (scintillating)
medium such as liquid argon has become more attractive \cite{gerda2}. 
\par
In this work, we demonstrate that CdZnTe room-temperature semiconductor
detectors also offer the option to be operated as a Heusser-type detector, in
liquid scintillator, in order to achieve ultra-low background levels for double-beta decay. 
Clearly, the possibility to operate a room-temperature setup significantly simplifies 
the arrangement of a large number of detector crystals in a tank and should
reduce overall costs.  
The great potential for extreme low background operation of
liquid scintillators alone has already been shown, for example, in the Borexino and
Kamland experiments \cite{borex}, \cite{kamland}.

\section{Experiment}
Our test setup is simple but sufficient for a proof-of-principle in a series
of planned experiments which are in progress. The first step was to identify
suitable liquid scintillator cocktails, i.e. mixtures that would permit
operation of electronics and high-voltage devices immersed within. No
selection with 
reference to optical and scintillating properties has been made at this stage. 
\par
\begin{figure}
\begin{center}
\includegraphics[width=14cm]{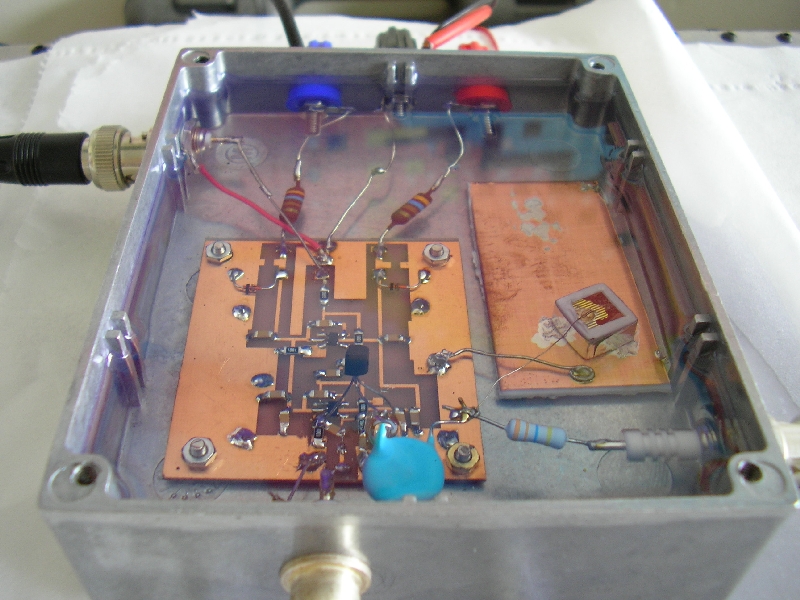}
\caption{Picture of the test setup, containing preamplifier and one of our
  Frisch collar crystals immersed in liquid scintillator.\label{photo}} 
\end{center}
\end{figure}

Crystals and preamplifier are
housed together in a standard diecast enclosure featuring 
connectors for preamplifier power, BNC output and High-voltage input (see
Fig.~\ref{photo}). A single
HV-power supply (Ortec 659, NIM module) delivers both polarities up to
5kV. Two linear DC power supplies deliver $\pm{}5$V to the preamplifier. The
output signal is connected directly to the data
acquisition system (DAQ) using a 50$\Omega$ BNC cable (RG58). The DAQ consists
of a 100 MHz sampling digital oscilloscope in a 3U PXI module from National
Instruments (NI PXI-5112) mounted in a PXI crate (NI PXI-1042) and is
controlled by an embedded controller PC (NI PXI-8186) running custom-made
LabView software for digital pulse acquisition. Pulses are streamed directly
to disk in binary format for maximum sampling speed when using radioactive
sources for detector calibration. 
\par 
The preamplifier is not a charge-sensitive amplification
system but a custom-made voltage amplifier. The motivation to try this type
initially, was to learn more about signal formation in the semiconductor,
i.e. specialise the analysis and measurement to pulse-shape, in contrast to
pulse-height. For a detailed discussion and circuit diagram, see
\cite{detpaper}. 
\par
Three coplanar grid detectors have been modified to work as
Frisch collar detectors \cite{collar}. All three are 1 cm cubes with
gold-plated electrodes, a full-area cathode and a coplanar grid anode structure. Five
faces (not the cathode face) are covered with insulating paint. Since we were
not interested in operating the grid, paint covering the anode was partly
removed (it dissolves easily in ethanol) and this area was used to contact
the anode with a wire attached by a generous drop of silver conductive
paint. The preamplifier is AC-coupled to the crystal anode which is biased
positively by the HV-power supply. The cathode is kept at ground
potential.  
\par
For the Frisch collar operation, each crystal was wrapped in two layers
of thin teflon tape, covering the full height, leaving out anode and cathode,
similar to devices fabricated in \cite{collar}. The teflon layer was then 
wrapped in a metal foil (aluminium kitchen foil worked best for us) and the
foil attached via a small 'lip' to the cathode (using silver
conductive paint). A crystal prepared thus was then mounted for operation on a
ground plane. We used a small copper-clad piece of printed-circuit board 
connected to the preamplifier ground, see Fig.~\ref{photo}. 
\par
We note that with respect to low background operation, some materials 
listed above are not
ideal, but can be replaced. The conducting foil could be high-purity, thin
copper foil, and the crystals could be contacted properly, either by bonding or
press-contact, again with a clean metal like copper, for example. Another
advantage of Frisch-collar detectors would be that they can be mounted in a
tightly packed array since physical contact of the metal shields around
crystals would be possible (all share a common ground). 
\par
\par
The metal foil
height determines the performance of the Frisch collar detector
\cite{collar}. We achieved best performance with 9 mm - 9.5 mm foil
height. Any higher shield results in stability problems when biasing the
anode since the grounded shield appears to be too close, particularly at the
cube corners. Again, for more details we refer to \cite{detpaper}.
\par
One major condition on the liquid scintillators was to respect health and
safety rules directly from the start, easing any potential practical
application. Fortunately, this is relatively straightforward since the advent
of 'safe' cocktails with low or no toxicity and no fire hazard. These types of
scintillators are commercially available from various companies. Our first
collection of 
scintillators was purchased from Perkin Elmer Life and
Analytical Sciences \cite{perkin} and National
Diagnostics \cite{national}. As stated above, an
assessment of optical and scintillating properties for these liquids is 
work-in-progress (for results on some liquids, see \cite{majewski}). We
identified four mixtures, each based on a different solvent,
which are viable, see Tab.~\ref{tab1}. We noticed that these viable cocktails
all seem to share a characteristic in common, namely that, according to
their manufacturers, they each specifically ``target organic samples'',
i.e. non-polar materials would be dissolved in the scintillator and
measured in typical life-science applications. It is assumed that a low
capacity of the cocktail to dissolve water
as the primary example of a polar sample material improves the resistivity of
the liquid.  

\section{Results}
Details of our data analysis procedure can be found in \cite{detpaper}. It is 
explained there how we achieved an energy resolution (FWHM) of 1.9\% at the
Cs-line (662 keV) for one of our Frisch-collar detectors (Det \#{}1) and just
above 2\% for the other two crystals. In this context it is interesting to
note that subsequently, we dedicated Det \#{}1 for the liquid scintillator
project, hence all results presented here can be seen as a follow-up of
\cite{detpaper} using the same detector. 
\par
The first remarkable change of properties occured after just a few days
immersion in 
the first liquid (Optiscint Hisafe). As was mentioned briefly in
\cite{detpaper}, the highest stable bias our detector could sustain at that
moment in time was 1kV. Measurements at higher bias showed increasing rates of
spurious pulses and deterioration of energy resolution. In liquid
scintillator however, the stability of detector operation improved to a degree
such 
that the stable operating bias was increased to 1.2kV with improved energy
resolution and total absence of spurious pulses. This statement is valid for
all scintillators tested and a collection of results can be seen in
Tab.~\ref{tab1}. Note that except for the Ecoscint liquid, all results improve
on the energy resolution obtained in air (1.9\%, see above and
\cite{detpaper}), even if only slightly.
\par
\begin{table}[htb]
\begin{center}
\vspace*{6pt}
\begin{tabular}{lll}
Scintillator  & Solvent & FWHM [\%{}] \\
Cocktail name & & (at 1200 Volt bias)  \\\hline
EcoScint O & PXE (Phenylxylylethane) &2.3\\
Optiscint Hisafe & DIN (Di-isopropylnaphtalene) &1.8\\
Mineral Oil &  & 1.8\\
Opti-Fluor O & LAB (Linear alkyl benzene) & 1.75\\\hline
\end{tabular}
\caption{Energy resolution (FWHM) results obtained in this work at 662 keV (Cs
  calibration source). Slightly inferior
  result were obtained for lower crystal bias. Measurements for
  each scintillator were repeated for three consecutive weeks before the
  cocktail was changed. Immediate measurements after such a liquid exchange
  consistently reproduced results obtained with the previous
  scintillator. Such a delay hints at the importance of the immediate liquid
  environment of the crystal, i.e. the dielectric in the small space between
  shield and teflon foil. The previous scintillator would initially still coat most of
  the surfaces and dilute only slowly into the new liquid.\label{tab1}}
\end{center}
\end{table}
\par
A second observation can be reported at this point, which might become crucial
for any low background
operation of CdZnTe crystals in liquid scintillator. Encouraged by the
stability and performance benefits of operating the detectors in liquid
scintillator, another test series was started. This time, the protective (and
unfortunately rather radioactive \cite{cobra}) paint on two crystals was
removed by wiping
with soft tissue and pure ethanol. Subsequently, as Frisch collar detectors in
air, it was confirmed that the crystals would not take high-voltage bias
beyond 400V without increased leakage current and an increasing rate of
spurious pulses. A subsequent immersion in the selected liquid scintillator
(based on energy resolution performance, see Tab.~\ref{tab1}),
Opti-Fluor O, 
for this long-term test, however, shows a remarkable recovery of
the CdZnTe. After 24 hours, the detector (Det \#{}1) takes up to a 1kV bias
and the following 
energy resolutions were obtained 
(at the Cs-137 line, 662 keV, energy resolution given as relative FWHM):
at 800V bias, 2.1\%, at 900V, 1.9\% and at 1kV,
2.0\%. After another full day in the scintillator, the detector takes the full
1.5kV bias and shows stable and consistent energy resolutions: at 800V, 2.2\%
and at bias values between 900V and 1.5kV: 1.9\% - 2.0\%.

\section{Conclusion}
We have demonstrated that a room-temperature Heusser-type detector made 
from CdZnTe semiconductors, immersed in a safe liquid scintillator cocktail is
a viable 
concept. Such a detector has the potential to achieve in a more practical and
cost-effective way what cryogenic versions \cite{gerda} set out to achieve,
i.e. a large mass, ultra-low background experiment to measure rare events like
double-beta decay. Additionally, prospects for low background operation of
CdZnTe crystals have been improved by demonstrating a stable operation of
clean, non-coated crystals directly in the insulating liquid
scintillator. Long-term tests, 
assessments of optical properties and collection of further suitable liquid
scintillators is work in progress. 

\begin{acknowledgments}
We wish to thank our collaborators on the COBRA project for providing the
CZT crystals which were used in this study.
\end{acknowledgments}

\end{document}